\documentclass{appolb}
\usepackage{graphicx}
\usepackage{epsfig}
\usepackage{amsmath}
\usepackage{amssymb}
\newcommand{\be}{\begin{equation}} 
\newcommand{\ee}{\end{equation}} 
\begin{document}
\title{Some aspects of final states and QCD evolution equations%
\thanks{Light Cone 2012, Krakow, Poland}%
}
\author{Krzysztof Kutak
\address{Instytut Fizyki J\c{a}drowej im. H. Niewodnicza\'nskiego\\
Radzikowskiego 152, 31-342 Krakow, Poland}
\\
}
\maketitle
\begin{abstract}
In order to study such effects like parton saturation in final states at the LHC one of the approaches is to combine physics of the BK and the CCFM
evolution equations. We report on recently obtained resummed form of the BK equation and nonlinear extension of the CCFM equation - the KGBJS equation.  
\end{abstract}


\PACS{PACS numbers come here}
  
\section{Introduction}
Quantum Chromodynamics (QCD) is a theory which 
is used to set up the initial conditions for the collisions at the LHC as well as to calculate properties of hadronic observables. 
In order to apply perturbative QCD to scattring process one needs to decompose the scross section into long distance part, called parton density, and a short distance part, called matrix element. In particular here we will focus on 
high energy factorization \cite{Catani:1990eg,Gribov:1984tu}. The evolution equations of high energy factorization resumme 
logarithms of energy accompanied by a strong coupling constant, i.e. terms proportional to $\alpha_s^n \ln^m s/s_0$. 
This framework applies if the total energy of a scattering process is much bigger than any other hard scale involved in a scattering process.\\
With the LHC one entered into a region of phase space where both the energy and
momentum transfers are high and partons form dense sysyem dense allowing in principle for parton saturation \cite{Albacete:2010pg,Dumitru:2010iy,Kutak:2012rf,Dusling:2012cg}. Recently a framework has been provided in \cite{Kutak:2011fu,Kutak:2012yr,Kutak:2012qk,vanHameren:2012uj,vanHameren:2012if} where both
dense systems and hard processes at high energies can be studied. 
\section{Exclusive form of the Balitsky-Kovchegov equation}
\begin{figure}[!t]
\centerline{\epsfig{file=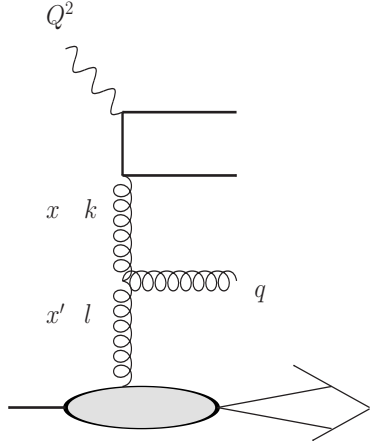,height=6cm,width=5cm}} 
\caption{{\it Plot explaining meaning of variables in BK and CCFM evolution equations.}}
\label{fig:Kinematics}
\end{figure}

At the leading order in $\ln 1/x$ the Balitsky-Kovchegov \cite{Balitsky:1995ub,Kovchegov:1999yj} equation for the WW gluon density in the momentum space is written as an integral equation reads \cite{Kutak:2011fu,Kutak:2012qk}: 
\begin{align}
\label{eq:fan1}
\Phi(x,k^2)&= \Phi_0(x,k^2)\\\nonumber
&+\overline\alpha_s\int_{x}^1 \frac{dz}{z}\int_0^{\infty}\frac{dl^2}{l^2}
\bigg[\frac{l^2\Phi(x/z,l^2)- k^2\Phi(x/z,k^2)}{|k^2-l^2|}+ \frac{
k^2\Phi(x/z,k)}{\sqrt{(4l^4+k^4)}}\bigg]\\\nonumber
&-\frac{\overline\alpha_s}{\pi R^2}\int_{x}^1 \frac{dz}{z}\Phi^2(x/z,k)
\nonumber
\end{align}
where the  lengths of transverse vectors lying in transversal plane to the collision axis are $k\equiv|{\bold k}|$, $l\equiv|{\bold l}|$ (${\bold k}$ is a 
vector sum of transversal momenta of emitted gluons during evolution), $z=x/x'$(see Fig. (\ref{fig:Kinematics}), $\overline\alpha_s=N_c\alpha_s/\pi$. The strength of the nonlinear term is controlled by the targets radius $R$. 
The linear term in eq. (\ref{eq:fan1}) can be linked to the process of creation of gluons while the nonlinear term
can be linked to fusion of gluons and therefore introduces gluon saturation effects.

The unintegrated gluon density obeying the high energy factorization theorem \cite{Catani:1990eg} is obtained from
 \cite{Braun:2000wr,Kutak:2003bd}: 
\be
{\cal F}_{BK}(x,k^2)=\frac{N_c}{\alpha_s \pi^2}k^2\nabla_k^2 \Phi(x,k^2)
\label{eq:glue}
\ee
where the angle independent Laplace operator is given by 
$\nabla_k^2=4\frac{\partial}{\partial k^2}k^2\frac{\partial}{\partial k^2}$.\\
As explained in \cite{Kutak:2011fu} this equation can be rewritten in a resummed form:
\be
\label{eq:bkexclusive}
\Phi(x,k^2)=\tilde \Phi^0(x,k^2)
+\overline\alpha_s\int_x^1d\,z\int\frac{d^2{\bf q}}{\pi q^2}\,
\theta(q^2-\mu^2)\frac{\Delta_R(z,k,\mu)}{z}\Bigg[\Phi(\frac{x}{z},|{\bf k} +{\bf q}|^2)\nonumber
\ee
\be
-\frac{1}{\pi R^2}q^2\delta(q^2-k^2)\,\Phi^2(\frac{x}{z},q^2)\Bigg].
\ee
where ${\bf q}={\bf l}-{\bf k}$ and $\Delta_R(z,k,\mu)\equiv\exp\left(-\overline\alpha_s\ln\frac{1}{z}\ln\frac{k^2}{\mu^2}\right)$ is a Regge form factor.\\ 
Eq. (\ref{eq:bkexclusive}) is a new form of the BK equation in which the resummed terms in a form of 
Regge form factor are the same for the linear and nonlinear part. 
This form will serve as a guiding equation to generalize the CCFM \cite{Ciafaloni:1987ur,Catani:1989sg,Catani:1989yc} equation to include nonlinear effects which allow for recombination of partons with constraint on angle of emission.
As has been shown in \cite{Kutak:2012qk} an analogous equation to (\ref{eq:bkexclusive}) can be written for the
high energy factorizable gluon density and it reads:
\begin{align}
\label{eq:bkres}
{\cal F}(x,k^2)=\tilde {\cal F}_0(x,k^2)+&\overline\alpha_s\int_{x/x_0}^1\frac{d\,z}{z}\Delta_R(z,k,\mu)\Bigg\{\int\frac{d^2q}{\pi q^2}\theta(q^2-\mu^2){\cal F}(\frac{x}{z},|k+q|^2)\\
&-\frac{\pi\alpha_s^2}{4N_cR^2}k^2\nabla_{k}^2\left[\int_{k^2}^{\infty}\frac{dl^2}{l^2}\ln\frac{l^2}{k^2}{\cal F}(x,l^2)\right]^2\Bigg\}\nonumber
\end{align}
\subsection{Nonlinear extension of the CCFM equation - the KGBJS equation }
As it has already been stated the motivation to extend the CCFM to account for nonlinearity is to be able to study the impact of saturation of partons on exclusive observables. There are indications that such effects might be significant in for instance 
production of charged particles at HERA \cite{Adloff:1996dy,Kutak:2008ed} or in forward production of di-jets \cite{Albacete:2010pg,Kutak:2012rf}.\\
The simplest nonlinear extension of CCFM has been recently proposed in \cite{Kutak:2011fu}
and its extension changes the interpretation
of the quantity for which the equation is written. It is not longer high energy factorizable gluon density but should be interpreted as the
dipole amplitude in momentum space or Weizsacker-Williams gluon density $\Phi$, denoted from now on by ${\mathcal E}$, which
besides $x$ and $k^2$ depends also on a hard scale $p$. The peculiar structure of the nonlinear term of the equation written below is motivated by the following 
requirements:
\begin{itemize}
\item the second argument of the $\mathcal{E}$ should be $k^2$ as motivated by the analogy to BK
\item the third argument should reflect locally the angular ordering
\end{itemize}
The KGBJS equation reads:
\be
\label{eq:final1}
\mathcal{E}(x, k^2, p)=\mathcal{E}_0(x,k^2,p)+\bar\alpha_s \int_x^1 dz
\int \frac{d^2\bar{\bf{q}}}{\pi \bar{q}^2} \,\theta (p - z\bar{q})
\Delta_s(p,z\bar{q})
\left ( \frac{\Delta_{ns}(z,k, q)}{z} + \frac{1}{1-z} \right )\times
\ee
\be
\Bigg[
\mathcal{E}\left(\frac{x}{z}, k^{'2}, \bar{q}\right)\\\nonumber
-\frac{1}{\pi R^2}\bar{q}^2\delta(\bar{q}^2-k^2)\,\mathcal{E}^2(\frac{x}{z},\bar{q}^2,\bar{q})\Bigg].
\ee
The momentum vector associated with $i$-th emitted gluon is
\be
q_i=\alpha_i\,p_P+\beta_i\,p_e+q_{t\,i}
\ee
The variable $p$ in (\ref{eq:final1}) is defined via $\bar{\xi} = p^2/(x^2s)$ where $\frac{1}{2}\ln(\bar{\xi})$ is a maximal rapidity which is determined by the kinematics of hard scattering,
$\sqrt{s}$ is the total energy of the collision and $k' = |\pmb{k} + (1-z)\bar{\pmb{q}}|$.
The momentum ${\bf\bar{q}}$ is the transverse rescaled momentum of the real gluon, and is related 
to ${\bf q}$ by $\bar{{\bf q}} = {\bf q}/(1-z)$ and $\bar q\equiv|{\bar{\bf q}}|$.\\ 
The form factor $\Delta_s$ screens the $1-z$ singularity
while form factor $\Delta_{ns}$ screens the $1/z$ singularity, in a similar form as the Regge form factor 
but also accounts for angular ordering:
\be
\Delta_{ns}(z,k,q)=\exp\left(-\alpha_s\ln\frac{1}{z}\ln\frac{k^2}{z q^2}\right).
\label{eq:nonsudakov}
\ee
where for the lowest value of $z q^2$ we use a cut-off $\mu$.

For recent numerical study and incorporation of higher order effects to equation (\ref{eq:final1}) we refer the Reader to \cite{Deak:2012mx}. 
Similarly as in case of the BK equation in order to obtain high energy factorizable unintegrated gluon density one applies relation (\ref{eq:glue}). 
The nonlinear term  in (\ref{eq:final1}),  apart from allowing for recombination 
of gluons might be understood as a way to introduce the decoherence into the emission pattern of gluons. 
This is because the gluon density is build up due to coherent gluon emission and since the nonlinear term 
comes with the negative sign it slows down the growth of gluon density and therefore it introduces the decoherence. We expect the nonlinear term to be
of main importance at low $x$ similarly as in case of the BK equation. In this limit it  will be of special interest to check whether in our formulation of the nonlinear 
extension of the CCFM equation one obtains an effect of saturation of the saturation
scale as observed in \cite{Avsar:2010ia} (for other formulations to include unitarity corrections in the CCFM evolution equation we refer the reader to \cite{Kutak:2008ed,Avsar:2009pf}). This effect is of great importance since it has a consequences for example for imposing a bound on amount of produced entropy from saturated part of 
gluon density as observed in \cite{Kutak:2011rb}. 
\section{Conclusions and outlook}
We reported on recently obtained new form of the BK equation written in a resummed form and on extension of the CCFM equation to account for nonlinearity. The obtained extension of the CCFM equation will be useful for studies of impact of saturation of gluons on exclusive observables.

\section{Acknowledgements}
Part of the research presented has been obtained in collaboration with Krzysztof Golec-Biernat, Stanislaw Jadach and Maciej Skrzypek. This research has been supported by grant LIDER/02/35/L-2/10/NCBiR/2011

\section{Bibliography}
\begin{footnotesize}

\end{footnotesize}

\end{document}